\documentstyle{article}
\title{ MATRIX MODELS FOR SPACETIME TOPODYNAMICS}

\author{R.R. Zapatrin,\\
Friedmann Laboratory for Theoretical Physics,\\
SPb UEF, Griboyedova 30/32,\\
191023, St-Petersburg, Russia}

\date{}

\begin{document}

\maketitle

\begin{abstract}
The machinery is suggested to describe the
varying spacetime topology on the level of its substitutes by
finite topological spaces.
\end{abstract}

\section*{Introduction}

The approximations of (or substitutes for) continuous spacetime by
finitary structures are studied in this paper. The results
presented furnish a framework in which one might express such ideas
as variable spacetime topology or, for instance, the topological
fluctuations on small scales. The paper is organized as follows.

In Section \ref{1} the coarse-graining procedure is described.
Being applied to a continuous manifold, it yields the so-called
{\it pattern space} \cite{prg} which, being finite or at most
countable set, may be thought of as topological space or,
equivalently, as a directed graph. The kinematics of the spacetime
topology is then addressed to that of the pattern space which
substitutes its continuous predecessor.

In Section \ref{2} the finitary counterpart of the supespace (in
Wheeler's sense) is introduced providing the arena for the
variation of the topology of pattern spaces. To construe it we use
the remarkable isomorphism between pattern spaces and finite
quasiorders. The latter, being subject of combinatorial studies,
are associated with certain finite-dimensional algebras
\cite{rota}. Thus, the study of the variety of finite topological
spaces (being, loosely speaking, discrete by its nature) is
replaced by dealing with finite-dimensional algebras whose matrix
representation is treated.

In Section \ref{3} the main topological features of pattern spaces
are formulated in algebraic terms.

In Section \ref{spat} the {\it spatialization procedure} is
suggested restoring points of the pattern spaces by given
finite-dimensional algebra. The ideas used in this procedure are
Stanley's techniques \cite{stanley} in algebraic combinatorics.

Now, possessing the algeraic means to capture the topological
features, we are interested in introducing finitary substitutes for
differential structures, to which the Section \ref{5} is
devoted.  The elements of the tensor calculus needed to introduce
the basic constituents of general relativity turn to be
successfully transplanted to pattern spaces and their matrix
representations.  The Einstein-Hilbert variational principle is
then rewritten in terms of matrix equations.

\medskip
\section{The coarse-graining procedure} \label{1}

\subsection{From manifolds to pattern spaces}

In the conventional general relativity, the spacetime manifold
consists of events. Whereas, from the operationalistic perspective
an individual event is an idealization of what can be directly
measured. Such idealization is
adequate within classical physics, but is unsatisfactory
from the operationalistic point of view. In quantum theory the
influence of a measuring apparatus on the object being
observed can not in principle be removed. We could expect the
metric of a quantized theory to be subject to fluctuations,
whereas the primary tool to separate individual events is just
the metric \cite{geroch1}. Thus a sort of smearing procedure
for events is to be imposed into the quantized theory of spacetime.

To introduce the procedure, recall the definition of the topology
$\tau$ of a manifold $M$. $\tau$ is nothing but a family of subsets
of $M$ declared {\sc open} and satisfying the following axioms:

\begin{itemize}
\item T1) $\emptyset ,X\in \tau$
\item T2) For any $A,B\in \tau A\cap B\in \tau$
\item T3) For any collection $\{A_{j}\}$, $A_{j}\in \tau\qquad
\bigcup_{j\in J}A_j\in \tau$, where $J$ is arbitrary index set,
$\cup , \cap $ are usual set union and intersection.
\end{itemize}

Thinking operationalistically, we can not have access to the
infinite number of all open sets, thus to capture the topology
of the manifold we consider its finite covering ${\cal F}$ by open
subsets which we believe to be homeomorphic to open balls in
${\bf R}^n$.

Supposed the covering ${\cal F}\subseteq\tau$ is closed under set
intersections, the spacetime manifold acquires the cellular
structure with respect to ${\cal F}$, so that the events belonging
to one cell are thought of as operationally undistinguishable.
Then, instead of considering the set $M$ of all events we can focus
on its finite subset $X\subseteq M$ such that each cell contains at
least one point of $X$.

For each $F\in {\cal F}$ (that is, $F\subseteq M$) consider
$F'=F\cap X$ (which is not empty by the choice of $X$) and treat
the collection ${\cal F}'=\{F'\mid F\in {\cal F}$ as the base of a
topology, denote it $\tau_X$ on $X$.

\medskip
\paragraph{Definition.} The finite topological space $(X, \tau_X)$
is called the {\sc pattern space} for the manifold $M$ with respect
to the covering ${\cal F}$.

\subsection{The graphs of pattern spaces}\label{graphs}

With each pattern space, being a finite topological space its Hasse
graph can be associated in the following way \cite{prg}. The
vertices of the graph are the points of $X$. Two points $x,y\in X$
are linked with the dart $x\to y$ if and only if the following
holds:

\begin{equation}
\forall A\in\tau_X\quad A\ni x\Rightarrow A\ni y
\label{darts}
\end{equation}

\noindent It can be verified directly that the obtained graph is
reflexive and transitive. Note that in general there may exist
points $x,y\in X$ such that $x\to y$ and $y\to x$, (see section
\ref{circ1}).

\paragraph{Lemma 1.} A subset $A\subseteq X$ is open if and only if
with each its point $a\in A$ it contains all the points $b\in X$
linked with $a$:

\begin{equation}
A\hbox{ is open }\qquad\Leftrightarrow\qquad\forall a\in A\quad
(\forall b\in X a\to b\Rightarrow b\in A)
\label{openg}
\end{equation}

\noindent{\it Proof} follows immediately from (\ref{darts}).
\paragraph{ Corollary.} A subset $B\subseteq X$ is closed if and
only if with each its point $b\in B$ it contains all $c\in X$ such
that $c\to b$:

\begin{equation}
B\hbox{ is closed }\qquad\Leftrightarrow\qquad \forall b\in B\quad
(\forall c\in X c\to b\Rightarrow c\in B)
\label{clog}
\end{equation}

\subsection{Example: a circle}\label{circ1}

Let $M=S^1$ be a circle: $M=\{e^{i\phi}{\rm mod}(2\pi)\}$. Consider
its covering ${\cal F}=\{F_1, F_2, F_3, F_4\}$ defined as:

\[
F_1  =  (\pi /4,3\pi /4) \qquad F_2  =  (-3\pi /4, -\pi /4)
\]
\[
F_3  =  (-3\pi /4,3\pi /4) \qquad F_4  =  (\pi /4, 7\pi /4)
\]

\noindent Let $X=\{0,\pi /2, \pi , -\pi /2, \pi /6\}$. Then the
graph of the appropriate pattern space is depicted on Fig.
\ref{circ1p}.

\unitlength 1mm
\begin{figure}[htb]
\begin{center}
\begin{picture}(70,30)
\put(0,0){\circle{2}} \put(2,0){$\pi$} %
\put(25,0){\circle{2}} \put(20,0){0} %
\put(0,25){\circle{2}} \put(2,25){$\pi/2$}
\put(25,25){\circle{2}} \put(27,25){$-\pi/2$}
\put(51,0){\circle{2}} \put(52,0){$\pi/6$} %
\put(0,1){\vector(0,1){23}}
\put(25,1){\vector(0,1){23}}
\put(1,1){\vector(1,1){23}}
\put(24,1){\vector(-1,1){23}}
\put(38,0){\vector(1,0){12}}
\put(38,0){\vector(-1,0){12}}
\end{picture}
\caption{ The pattern circle.}
\label{circ1p}
\end{center}
\end{figure}
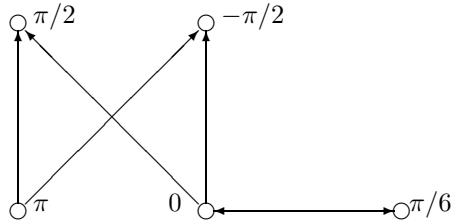

\noindent In the examples presented below the transitive darts will
often be omitted.

\subsection{Example: the real line}\label{line1}
\medskip

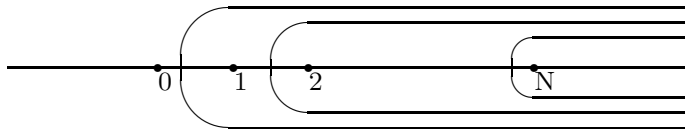
\begin{figure}[htb]
{\unitlength 1mm
\begin{picture}(90,28)
{\thicklines
\put(0,20){\line(1,0){90}}
} 
\put(20,20){\circle*{1}}
\put(20,17){\sc 0}
\put(30,20){\circle*{1}}
\put(30,17){\sc 1}
\put(40,20){\circle*{1}}
\put(40,17){\sc 2}
\put(70,20){\circle*{1}}
\put(70,17){\sc N}
\put(30,20){\oval(14,16)[l]}
\put(40,20){\oval(10,12)[l]}
\put(70,20){\oval(6,8)[l]}
\put(30,28){\line(1,0){60}}
\put(40,26){\line(1,0){50}}
\put(70,24){\line(1,0){20}}
\put(30,12){\line(1,0){60}}
\put(40,14){\line(1,0){50}}
\put(70,16){\line(1,0){20}}
\end{picture}
}
\caption{The covering of the real line.}
\label{line1fig}
\end{figure}

Let $M={\bf R}^1$. Fix up a positive integer $N$ and consider
the covering ${\cal F}=\{F_i\mid i=0,\ldots N\}$ (Fig.
\ref{line1fig}) with

\begin{eqnarray}
F_0 & = & {\bf R}^1\\
F_i & = & (i-1/2, +\infty )
\label{efi}
\end{eqnarray}


\begin{figure}[htb]
{\unitlength 1mm
\thicklines
\begin{picture}(90,5)
\put(10,10){\circle*{1}}
\put(10,1){\sc 0}
\put(11,0){\vector(1,0){13}}
\put(25,0){\circle*{1}}
\put(25,1){\sc 1}
\put(26,0){\vector(1,0){13}}
\put(40,0){\circle*{1}}
\put(40,1){\sc 2}
\put(41,0){\vector(1,0){13}}
\put(56,0){\ldots}
\put(61,0){\vector(1,0){13}}
\put(75,0){\circle*{1}}
\put(75,1){\sc N}
\end{picture}
} 
\caption{The pattern real line ${\bf R}^1$}
\label{line1p}
\end{figure}
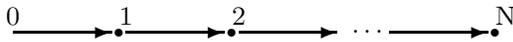

Let $X=\{0,1,\ldots ,N\}$, then the Hasse graph of $X$ is
presented on Fig. \ref{line1p}.

\medskip
\section{ Algebraic superspace}\label{2}
\subsection{ Quasiorders and partial orders}

As it was already established in \ref{graphs}, each pattern space
can be associated with a reflexive and transitive directed graph.
When such a graph is set up, we may consider its darts as pointing
out a relation between the points of $X$, denote it also $\to$.
This relation has the properties:

\[ \forall x\in X\qquad  x\to x \]
\begin{equation}
\forall x,y\in X \qquad x\to y\hbox{ and }y\to z\hbox{ imply }x\to z
\label{qo}
\end{equation}

A relation on an arbitrary set having the properties (\ref{qo}) is
called {\sc quasiorder}. When a quasiorder is antisymmetric:

\begin{equation}
\forall x,y\in X\quad x\to y\hbox{ and }y\to x\hbox{  imply  }x=z
\label{po}
\end{equation}

\noindent the relation $\to$ is called {\sc partial order}.
\medskip

\subsection{ Incidence algebras}

In the case when $M$ is a compact manifold, there is the algebra
${\cal A}=C^\infty (M)$ of all smooth functions on $M$. ${\cal A}$
can be treated as the algebraic substitute of $M$ in that sense
that, given ${\cal A}$ considered algebra ({\it i.e.} linear space
with associative product operation), there exist the algebraic
techniques (the Gel'fand procedure) which restore the points of $M$
together with its topology. In the case when $X$ is a finite
topological space, the attempts to consider even a broader algebra
of continuous functions $X\to {\bf C}$ fails since the structure of
such algebra captures only the number of connected components of
$X$ and nothing more \cite{prg}. Although, if we treat $X$ as
quasiordered set, we can broaden a well-known algebraic scheme from
combinatorics, namely, that of incidence algebra \cite{rota},
slightly generalized to pattern spaces (being quasiordered sets, in
general).

\paragraph{Definition.} For a quasiordered set $X$ define its
{\sc incidence algebra} ${\cal A}_X$, or simply ${\cal A}$ if no ambiguity
occurs, as the collection of all complex-valued functions of two
arguments vanishing on non-comparable pairs:

\begin{equation}
{\cal A}=\{a:X\times X\to {\bf C}\mid a(x,y)\neq 0\Rightarrow x\to
y\}\label{a}
\end{equation}

To make the defined linear space ${\cal A}$ algebra we define the
product of two elements $a,b\in {\cal A}$ as:

\begin{equation}
ab(x,y)=\sum_{z:x\to z\to y}^{} a(x,z)b(z,y)
\label{aprod}
\end{equation}

\noindent It can be proved that the so-defined product operation is
associative \cite{rota}. Since the set $X$ is finite, the algebra
${\cal A}$ is finite-dimensional associative (but not commutative, in
general) algebra over ${\bf C}$.

Now let us clear out the meaning of the elements of ${\cal A}$.
Let $a\in {\cal A}$ and $x,y$ be two points of $X$. If they are not
linked by a dart then, according to (\ref{a}), the value $a(x,y)$
always vanishes. So, $a(x,y)$ can be thought of as an assignment of
weights (or, in other words, transition amplitudes) to the darts
of the graph $X$. In these terms the product (\ref{aprod}) has the
following interpretation. Let $c=ab$, then $c(x,y)$ is the sum of
the amplitudes of all allowed two-step transitions, the first step
being ruled by $a$ and the second by $b$ (Fig. \ref{trans}),

\unitlength 1mm
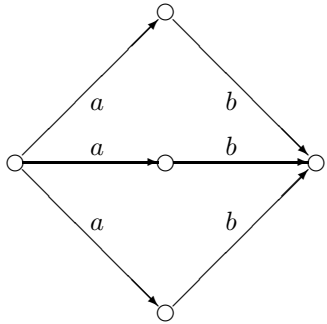
\begin{figure}[htb]
\begin{center}
\begin{picture}(70,42)
\put(20,0){\circle{2}}
\put(0,20){\circle{2}}
\put(20,20){\circle{2}}
\put(40,20){\circle{2}}
\put(20,40){\circle{2}}
\put(10,11){$a$}
\put(10,21){$a$}
\put(10,27){$a$}
\put(28,11){$b$}
\put(28,21){$b$}
\put(28,27){$b$}
\put(1,19){\vector(1,-1){18}}
\put(1,20){\vector(1,0){18}}
\put(1,21){\vector(1,1){18}}
\put(21,39){\vector(1,-1){18}}
\put(21,20){\vector(1,0){18}}
\put(21,1){\vector(1,1){18}}
\end{picture}
\caption{ Allowed transitions on pattern space.}
\label{trans}
\end{center}
\end{figure}

\noindent while
the element $c(x,y)$ of the multiple product $c=a_1\ldots a_n$
looks similar to the Feynman sum over all paths from $x$ to $y$
allowed by the graph $X$ of the length $n$ and the closest
physical counterpart of the elements of the incidence algebra are
$S$-matrices.

So, the transition from pattern spaces to algebras is described.
The inverse procedure of "spatialization" will be described below
in the Section \ref{spat}.

\subsection{The standard matrix representation of incidence
algebras}\label{157}

Given the incidence algebra of a pattern space $X$, its standard
matrix representation is obtained by choosing the basis of ${\cal A}$
consisting of the elements of the form ${\bf e}_{ab}$, where $ab$
range over all ordered pairs $a\to b$ of elements of $X$, defined
as:

\begin{equation}
{\bf e}_{ab}(x,y)= \cases{
1 & $x=a$ and $y=b$ (provided $a\to b$)\cr
0 & otherwise}
\label{eab}
\end{equation}

\noindent We can also extend the ranging to {\em all} pairs of
elements of $X$ by putting ${\bf e}_{ab}\equiv 0$ for $a\not\to b$.
Then the product (\ref{aprod}) reads:

\begin{equation}
{\bf e}_{ab}{\bf e}_{cd} = \delta_{bc}{\bf e}_{ad}
\label{mprod}
\end{equation}

\noindent With each $a\in {\cal A}$ the following $N\times
N$-matrix ($N$ being the cardinality of $X$) is associated:

\[
a\mapsto a_{ik} = a(x_i,x_k)
\]

\noindent Let $I$ be the incidence matrix of the graph $X$, that is

\begin{equation}
I_{ik} = \cases{
1 & $x_i\to x_k$\cr
0 & otherwise}
\label{inc}
\end{equation}

\noindent then the elements of ${\cal A}$ are represented as the
matrices having the following property:

\begin{equation}
\forall i,k \quad a_{ik}I_{ik} = a_{ik}\qquad\hbox{no sum over }i,k
\label{ai}
\end{equation}

\noindent The product $c=ab$ of two elements is the usual matrix
product:

\[
c_{ik} = c(x_i,x_k) = \sum_{i\to l\to k}^{}a(x_i,x_l)b(x_l,x_k) =
\sum_{forall l}^{} a_{il}b_{lk}
\]

That means, we have so embedded ${\cal A}$ into the full matrix algebra
${\cal M}_N({\bf C})$, that ${\cal A}$ is represented by the set of {\em
all} matrices satisfying (\ref{ai}). So, to specify an incidence
algebra in the standard representation we have to fix the template
matrix $I_{ik}$ (\ref{inc}). We can always re-enumerate the
elements of $X$ to make the template $I_{ik}$ {\em
upper-block-triangular} matrix with the blocks corresponding to
cliques. In particular, when $X$ is partially ordered, each
clique contains exactly one element of $X$, and the incidence
matrix $I$ is upper triangular.

\subsection{Examples}
\medskip

Return to the examples of Section \ref{1}. The first example was
the circe simulated by the pattern space $X$ (Fig. \ref{circ1p})
having the following incidence matrix:

\[
I_{\rm circle}=\left(\begin{array}{ccccc}
1 & 0 & 0 & 1 & 1\\
0 & 1 & 1 & 1 & 1\\
0 & 1 & 1 & 1 & 1\\
0 & 0 & 0 & 1 & 0\\
0 & 0 & 0 & 0 & 1
\end{array}\right)
\]

The standard matrix representation of the real line yields the
algebra $T_N$ of all upper triangular matrices with the incidence
matrix

\[
I_{\rm line}=\left(\begin{array}{ccccc}
1 & 1 & 1 & \ldots & 1\\
0 & 1 & 1 & \ldots & 1\\
0 & 0 & 1 & \ldots & 1\\
0 & 0 & 0 & \ldots & 1\\
\ldots & \ldots & \ldots & \ldots & \ldots\\
0 & 0 & 0 & \ldots & 1
\end{array}\right)
\]

\subsection{The algebraic superspace}

Now we are in a position to introduce the arena for the future
topodynamics of pattern spaces. If we fix up the cardinality $N$,
we already know that any pattern space of this cardinality can be
isomorphically \cite{rota} restored by its incidence algebra
which, in turn, can be represented by $N\times N$ matrices. So, the
hazy question of what is the room for all topologies has the natural
solution in finitary case: we can consider the space ${\cal
M}={\rm Mat}_N({\bf C})$ of {\em all} $N\times N$ matrices. From
now on this space will be referred to as {\sc finitary algebraic
superspace}.

As it will be shown in section \ref{spat}, any subalgebra of ${\cal
M}$ gives rise to a finite topological space. So, if we treat
${\cal M}$ as state space, the topologies are associated with the
subspaces of ${\cal M}$ with the only discrepancy with the
conventional quantum mechanical approach that these subspaces are
closed in a specific algebraic sense. Namely, for any subspace
${\cal A}\subseteq {\cal M}$ its closure is build as the algebraic
hull of ${\cal A}$, that is, the intersection of all subalgebras of
${\cal M}$ containing ${\cal A}$.

\medskip
\section{Topological features and constructions in algebraic
terms}\label{3}

\subsection{Connectedness}
\medskip

A topological space $X$ is called {\sc connected} if it contains no
proper subsets being both closed and open, and {\sc linearly
connected} if each pair $x,y$ of its points can be connected by a
path (a continuous mapping $p:[0,1]\to X$ such that $p(0)=x$ and
$p(1)=y$). For finite topological spaces the notions of
connectedness and linear connectedness coincide.

\paragraph{Definitions.} Given a directed graph $X$, its {\sc
underlying graph} ${\bf UN}X$ is the undirected graph obtained from
$X$ by forgetting the direction of all its darts. An {\sc
undirected path} in $X$ connecting $x,y\in X$ is the sequence
$x=x_0, x_1,\ldots ,x_n=y$ of vertices of $X$ such that each pair
$(x_i,x_{i+1})$ is linked by an arc of the underlying graph ${\bf
UN}X$.

\paragraph{Statement.} Let $x,y\in X$. Then they can be connected
by a path (in the topological sense) if and only if they can be
connected by an undirected path defined above.

\noindent{\it Proof.} See \cite{prg}.

\subsection{Disjoint sums}

Let $X_1, X_2$ be two pattern spaces such that $X_1\cup
X_2=\emptyset$ and $X:=X_1\cap X_2$ be their disjoint sum. Denote
by ${\cal A}={\cal A}(X)$, ${\cal A}_1={\cal A}(X_1)$, ${\cal
A}_2={\cal A}(X_2)$ the appropriate incidence algebras.

\paragraph{Statement.} ${\cal A}$ is the direct sum:

\begin{equation}
{\cal A} = {\cal A}_1\bigoplus {\cal A}_2
\label{dirsum}
\end{equation}

\noindent{\it Proof}. Since $X_1\cup X_2=\emptyset$, both $X_1$ and
$X_2$ are clopen subsets of $X$, hence, according to (\ref{openg}),
no pair of points $x^1_i\in X_1$ and $x^2_k\in X_2$ can be
connected by a dart of $X$. Therefore for any any $a\in {\cal A}$
we have $a(x^1_i,x^2_k) = a(x^1_k,x^2_i) = 0$. Consider the
projections $\pi_\alpha :{\cal A}\to {\cal A}_\alpha$, $\alpha =
1,2$ defined as follows:

\[
(\pi_\alpha a)(x^\alpha_i,x^\alpha_k) = a(x^\alpha_i,x^\alpha_k)
\]

\noindent and the injection $i_1 :{\cal A}_1\to {\cal A}$:

\[ (i_1 a_1)(x^1_i, x^1_k) = a_1 (x^1_i, x^1_k) \]

\[ (i_1 a_1)(x^2_i,y) = (i_1a_1)(y,x^2_i) = 0\qquad \forall
y\in X \]

Clearly $i_1\pi_1i_1\pi_1 = i_1\pi_1$ and $\pi_1i_1a_1 = a_1$ for
all $a_1\in {\cal A}_1$, hence ${\cal A}_1$ is the direct summand
of ${\cal A}$, as well as ${\cal A}_2$ which is proved in the same
way.

\subsection{Boolean machinery}
\medskip

In \cite{prg} the {\em Boolean machinery} to transform pattern
spaces was suggested. A pattern space was treated as directed graph
and two basic operations of {\em stretching} and {\em cutting}
darts were introduced. This allowed to consider the stepwise
changing of the topology of the pattern space. Let us translate
these operations into the language of incidence algebras. Begin
with the stretching operator.
\medskip

Let $X$ be a pattern space and $a,b$ ($a\neq b$) be a pair of its
vertices. The {\sc stretching operator} $S^{ab}$ stretches the dart
from $a$ to $b$ (in particular, does nothing if the dart already
exists). The result of this only stretching may yield a
non-transitive (\ref{qo}) graph, so, to stay within pattern spaces
we have to add the lacking darts to make it transitive. In the
language of incidence algebras this procedure looks as follows.  We
have ${\cal A}_X\subseteq {\cal M}$ in the standard representation
(\ref{eab}). Applying $S^{ab}$ means adding ${\bf e}_{ab}$ to the
set of basis vectors and then we form the linear span of $({\cal
A}_X \cup\{{\bf e}_{ab}\}$. The result may not be the subalgebra of
${\cal M}$, so, to render it subalgebra we have to consider its
algebraic span, which results in adding the basis elemnts ${\bf
e}_{a'b}$ for each ${\bf e}_{a'a}\neq 0$ and ${\bf e}_{ab'}$ for
each ${\bf e}_{bb'}\neq 0$.

The {\sc cutting operator} $C^{ab}$ removes the element ${\bf e}_{ab}$
from the the basis, and then the algebraic hull is formed. It may
happen that the dart $a\to b$ being composite is unremovable and we
return to the same algebra, and, hence, to the same pattern space.

\medskip
\section{The spatialization procedure}\label{spat}

This section describes the procedure reverse to that described in
the section \ref{2}. Namely, the suggested spatialization procedure
having a finite-dimensional algebra on its input, manufactures a
quasiordered set. Being applied to an incidence algebra ${\cal A}_X$ of
a quasiordered space $X$, it yields the initial space $X$ (up to a
graph isomorphism).

It is assumed that the Reader of this section is familiar with the
basic notions of the theory of associative algebras such as ideal,
radical, semisimplicity and so on, and I will use these terms
without defining them. Although, it seems appropriate to introduce
the necessary definitions from the theory of partially ordered
sets.

\subsection{Interlude on partial and quasiorders}

Let $(Y,\to )$ be a quasiordered set (\ref{qo}). Define the
relation $\sim$ on $Y$

\[
x\sim y\qquad\Leftrightarrow\qquad x\to y\hbox{ and }y\to x
\]

\noindent being equivalence on $Y$, and consider the quotient set
$X=Y/\sim$. Then $X$ is the partially ordered set \cite{lt}.

When $Y$ is a pattern space, the transition from $Y$ to $X$ has the
following meaning: $X$ is obtained from $Y$ by smashing cliques to
points. Contemplating this procedure we see that $X$ may also be
treated as the subgraph obtained from $Y$ by deleting all but one
'redundant' vertices with adjacent (both incoming and outgoing)
darts.

Now let us study how the relation between the quasiorders and
associated partial orders looks in terms of incidence algebras. Let
${\cal A} = {\cal A}(Y)$ and ${\cal A}' = {\cal A}(Y/\sim )$.  Then
${\cal A}'$ is the subalgebra of ${\cal A}$. In the standard matrix
representation (\ref{eab}) ${\cal A}'$ is obtained as follows:

\begin{enumerate}
\item Select the set $X_R\subseteq X$ of redundant vertices (say,
by checking out the identical rows of the incidence matrix $I$
(\ref{inc})
\item Select the set $E_R$ of redundant basis elements (\ref{eab}):
\begin{equation}
{\bf e}_{ab}\in E_R\qquad\Leftrightarrow\qquad a\in X_R\hbox{ or }b\in
X_R
\label{redeab}
\end{equation}
\item Delete the elements of $E_R$ from the basis, then
\begin{equation}
E_X = \{\hbox{the basis of }{\cal A}_Y\}\setminus E_R
\label{eabp}
\end{equation}
\noindent is the basis of the incidence algebra ${\cal A}_X$.
\end{enumerate}
\medskip

We shall also consider the inverse procedure of {\sc expanding} a
partially set $X$ to a quasiorder $Y$. To each point of $x\in X$ a
positive integer $n_x$ is assigned (which can be thought of as a
sort of inner dimension --- a room for gauge transformations). Then
each $x$ is replaced by its $n_x$ copies linked between each other
by two-sided darts and having all the incoming and outgoing darts
the same as $x$.

So, given a quasiordered set $Y$, we can always represent it as the
partially ordered set $x$ of its cliques equipped with the
additional structure: to each $x\in X$ an integer $n_x\ge 1$
thought of as the cardinality of appropriate clique is assigned:
\begin{equation}
Y = (X, n_x)
\label{qopo}
\end{equation}

\subsection{The spatialization procedure.}

Now let us explicitly describe the construction which will build
pattern spaces by given finite-dimensional algebras. Let ${\cal A}$
be a subalgebra of the full matrix algebra ${\rm Mat}(n,{\bf C})$.
Denote by ${\cal R}$ the radical of the algebra ${\cal A}$. To
build the pattern space associated with ${\cal A}$ the following is
to be performed.

\begin{itemize}
\item {\bf Step 1. Creating cliques.} Form the quotient ${\cal A}'
= {\cal A}/{\cal R}$ (being always semisimple since ${\cal R}$ is
the radical of ${\cal A}$). Denote by ${\cal K}$ the center of
${\cal A}'$:

\[
{\cal K} = {\rm Center}({\cal A}/{\cal R})
\]

Then define the set $X$ of {\sc cliques} as the set of all
characters (i.e. linear multiplicative functionals) on ${\cal K}$:

\[
X := \chi ({\cal K})
\]

\item {\bf Step 2. Assigning cardinality to cliques.} Since the
algebra ${\cal A}$ is semisimple it is the direct sum of simple algebras
and the set $X$ labels its simple components ${\cal A}'_x$:

\[
{\cal A}' = \oplus_{x\in X}{\cal A}'_x
\]

\noindent Each ${\cal A}'_x$ being simple finite-dimensional algebra has
the exact representation as the algebra of all $n_x\times n_x$
matrices. Assign this number $n_x$ to each $x\in X$ and call it the
{\sc cardinality} of the clique $x$.

\item {\bf Step 3. Stretching the darts.} For each character $x\in
X$ consider its annihilating subset ${\cal K}_x\subseteq {\cal
K}\subseteq {\cal A}'$ and span on it the two-sided ideal ${\cal
N}'_x$ in ${\cal A}'$:

\[
{\cal N}'_x := {\cal A}'{\cal K}_x{\cal A}'
\]

\noindent and consider its preimage ${\cal N}_x$ with respect to the
canonical projection $\pi :{\cal A}\to {\cal A}' = {\cal A}/{\cal R}$:

\[ {\cal N}_x := \pi^{-1}({\cal N}'_x) \]

\noindent being the ideal in {\cal A}.

For each pair $x,y\in X$ , $x\neq y$ form two linear subspaces of
${\cal A}$: ${\cal N}_x\cap {\cal N}_y$ and ${\cal N}_x\cdot {\cal
N}_y$ and consider the quotient linear space:

\begin{equation}
Q(x,y) := {\cal N}_x\cap {\cal N}_y/{\cal N}_x\cdot {\cal N}_y
\label{qxy}
\end{equation}

Then, if and only if $Q(x,y)\neq 0$ stretch the dart $x\to y$.
\end{itemize}

\paragraph{Remark.} The last step is based
on the well known construction called scheme of a
finite-dimensional algebra.

When (\ref{qxy}) is checked for all $x,y$, the non-transitive
predecessor of the partially ordered set $X$ is obtained. To have
$X$, form the transitive closure of $X$:

\[
{\rm darts}X := \{(x,x)\}_{x\in X}\cup \{(x,z)\mid \exists x = y_0,
\ldots ,y_n = z Q(y_i,y_{i+1})\neq 0\}
\]

So, the pattern space $Y = (X,n_x)$ (\ref{qopo}) is completely
built. In the sequel denote the quasiordered set furnished by the
spatialization procedure applied to the algebra ${\cal A}$ by ${\rm
spat}$:

\begin{equation}
Y = {\rm spat}{\cal A}
\label{dspat}
\end{equation}

\paragraph{Remark.} Being applied to the incidence algebra of a
quasiorder $Y$, this procedure restores $Y$ up to an isomorphism of
quasiorders, as it follows from the Stanley's theorem
\cite{stanley}.
\medskip

\medskip
\section{Differential geometry in algebraic superspace } \label{5}

It was shown in the previous section how, starting from a
particular algebra ${\cal A}$ to extract from it something which
may be interpreted as substitute of spacetime. Although, the
question arises where this algebra ${\cal A}$ can be taken from. As
it was already outlined in section \ref{2}, the variety of this
algebras can be placed into an algebraic superspace whose role in
the case of fixed cardinality $N$ of the pattern space we expect to
be furnished, may be played by the full matrix algebra ${\rm
Mat}(N,{\bf C})$. In this section I am going to formulate the
finitary counterparts of differential structures in which the basic
notions of general relativity are formulated. For the overview of
the mathematical problems related with such reinterpretation the
Reader is referred to \cite{ps}.

\subsection{Basic algebras.}

Loosely speaking, the notion of basic algebra is the non-commutative
generalization of Einstein algebras suggested by Geroch. His main
observation was \cite{gerochea} that, building general relativity,
the notion of the spacetime manifold $M$ is essentially used only
once: to define the algebra ${\cal A} = C^\infty{}(M)$ of all smooth
functions on $M$. All the forthcoming notions can be then
reinterpreted in mere terms of ${\cal A}$. For instance, {\sc vector
fields} are defined as derivations of ${\cal A}$, that is linear
mappings $v:{\cal A}\to {\cal A}$ enjoying the Leibniz rule:

\begin{equation}
v(ab) = va\cdot b + a\cdot vb
\label{leib}
\end{equation}

\noindent and so on.

So, let ${\cal S}$ be an algebraic superspace, that is, a finite
dimensional algebra. Choose the basis $\{E_0,E_1,\ldots\}$ of
${\cal S}$:

\[ {\cal A} = {\rm span}\{E_0, E_k\mid k = 1,\ldots\} \]

\noindent so that

\[ E_0  =  1 \qquad{\rm Tr}E_k  =  0 \]

\noindent where $1$ is the unit element of ${\cal A}$ (the meaning
of this requirement will be clarified below).

The product in ${\cal S}$ can be defined in terms of structural
constants:

\begin{equation}
E_0E_i = E_iE_0 = E_i \qquad\hbox{ ; }\qquad
E_iE_k = P^l_{ik}E_l
\label{eprod}
\end{equation}

\noindent so that the associativity condition $(E_iE_k)E_l = E_i(E_kE_l)$
holds:

\begin{equation}
P^m_{ik}P^n_{ml} = P^n_{im}P^m_{kl}
\label{ass}
\end{equation}

Since ${\cal S}$ is in general non-commutative, the commutators of basis
elements do not vanish. Denote them $L^l{}_{ik}$:

\[
E_iE_k  =  E_iE_k - E_kE_i  =  2L^l{}_{ik}E_l
\]
\begin{equation}
L^l{}_{ik} = \frac{1}{2}(P^l_{ik} - P^l_{ki}) = -L^l{}_{ik}
\label{liec}
\end{equation}

\subsection{Vectors. }

The vectors are the derivations of the algebra ${\cal S}$. In the case
when ${\cal S} = {\rm Mat}(n,{\bf C})$ all derivatives are exhausted by
inner ones. Moreover, they are in 1-1 correspondence with
zero-trace matrices. So, if we choose the basis $E_1,\ldots$
elements of ${\cal S}$ to be of zero trace, they can serve as the basis
for the space $V$ of all vectors as well:

\begin{equation}
V = {\rm span}\{E_k\mid k = 1,\ldots\}
\label{lieb}
\end{equation}

\noindent The space $V$ is the Lie algebra with respect to matrix
commutation, therefore the constants $L^l{}_{ik}$ (\ref{liec}) are
the Lie constants for $V$. Note that we always have the embedding
$E_k\to E_k$ of $V$ into ${\cal S}$ as that of linear spaces.

Being finite-dimensional Lie algebra, $V$ possesses two canonical
forms: the trace of the Lie constants

\begin{equation}
L_i = L^m{}_{mi}
\label{li}
\end{equation}

\noindent and the Killing form

\begin{equation}
K_{ij} = L^m{}_{ni}L^n{}_{mj}
\label{kij}
\end{equation}

\paragraph{Remark.} In the case of the standard basis $\{{\bf
e}_{ab}\}$ (\ref{eab}) of matrix units, the interpretation of
the elements of $V$ as vector fields becomes very transparent.
Namely, each derivation is associated with the assignment of a
weight to each dart of the graph of pattern space.

\subsection{ Connection, curvature and all that.}

Since the space $V$ plays the role of "tangent bundle", the
{\sc connection} can be defined as a linear mapping $D:V\times V\to
V$ being derivation with respect to the second argument.

\begin{equation}
D(E_i,E_kE_l) = D(E_i,E_k)E_l + E_kD(E_i,E_l)
\label{dleib}
\end{equation}

\noindent A particular choice of the connection $D$ can be set up by
defining the set of appropriate structural constants

\begin{equation}
D(E_i,E_k) = D^m{}_{ik}E_m
\label{dik}
\end{equation}

\noindent then (\ref{dleib}) reads:

\[
D^n{}_{im}P^n{}_{kl} =
D^m{}_{ik}P^n{}_{ml} + D^m{}_{il}P^n{}_{km}
\]

\noindent Now, when the connection is defined, introduce the {\sc
torsion} associated with the connection $D$:

\[
T(E_i,E_k) := D(E_i,E_k) - D(E_k,E_i) - [E_i,E_k]
\]

\noindent In the sequel we shall be interested in torsion-free
connections: $T(E_i,E_k) = 0$. It can be expressed as the
following relations between the structural constants: $D^l{}_{ik} -
D^l{}_{ki} -2L^l{}_{ik} = 0$, or, in a more suitable form:

\begin{equation}
D^l{}_{ki} = D^l{}_{ik} - 2L^l{}_{ik}
\label{t0}
\end{equation}

In the non-commutative environment we can keep using the standard
definition of Riemann and Ricci forms (see, e.g. \cite{he}):

\begin{equation}
R(X,Y)Z = D_X(D_YZ) - D_Y(D_XZ) - D_{[X,Y]}Z
\label{riemg}
\end{equation}

\noindent Substitute basis vectors $E_i,E_k,E_l$ into
({\ref{riemg}) and decompose the left side over the basis denoting
$R(E_i,E_k)E_l$ by $R^n{}_{ikl}E_n$, then these coefficients are
expressed via Christoffel symbols $D^l{}_{ik}$ (\ref{dik}) as
follows:

\begin{equation}
R^n{}_{ikl} = D^n{}_{im}D^m{}_{kl} -
D^n{}_{km}D^m{}_{il} - 2D^n{}_{ml}L^m{}_{ik}
\label{riemt}
\end{equation}

The Ricci tensor is the trace of the Riemann one, and we can obtain
its coefficients by contracting (\ref{riemt}) over the pair of
indices $n$--$k$:

\[ R_{il} = R^k{}_{ikl} =
D^k{}_{im}D^m{}_{kl} -
D^k{}_{km}D^m{}_{il} - 2D^k{}_{ml}L^m{}_{ik}
\]

\noindent which yields after replacing the summation indices in the
first term and taking (\ref{t0}) into account:

\begin{equation}
R_{il} =
D^k{}_{mi}D^m{}_{kl} - D^k{}_{km}D^m{}_{il}
\label{ric0}
\end{equation}

\subsection{The metric and concertedness condition.}

The metric tensor can be defined as a Hermitian form on $V$ taking
the values in ${\cal S}$. So, its most general form is:

\begin{equation}
g(E_i,E_k) = g_{ik}E_0 + g^m{}_{ik}E_m
\label{g0}
\end{equation}

\noindent with $g_{ik} = \overline g_{ki}$ and $g^m_{ik} =
\overline g^m_{ki}$, where the bar means usual complex conjugation.
The requirement of nondegeneracy of the metric can be formulated
here in different ways. We shall impose it in the following form:

\begin{equation}
{\rm det}g_{ik} \neq 0 \quad\hbox{ and }\quad
{\rm det}g^m_{ik} \neq 0 \qquad \forall m
\label{ndg}
\end{equation}

By {\sc concertedness condition} I mean the analog of the
well-known classical requirement $\nabla_X (g(Y,Z)) = g(\nabla_X
Y,Z) + g(Y,\nabla_X Z)$ which can be rewritten as follows:

\begin{equation}
D(E_i,g(E_k,E_l)) =
g(D(E_i,E_k),E_l) + g(E_k,D(E_i,E_l))
\label{conc}
\end{equation}

\noindent Substituting (\ref{dik},\ref{g0}) to (\ref{conc}) and
assuming the covariant derivative of $E_0$ to be zero, we obtain:

\begin{equation}
\begin{array}{rcl}
0                    & = & g_{ml}D^m{}_{ik} + g_{km}D^m{}_{il} \cr
D^n{}_{im}g^m{}_{kl} & = & g^n_{ml}D^m{}_{ik} + g^n_{km}D^m{}_{il}
\end{array}
\label{c}
\end{equation}

\noindent and consider (\ref{c}) as equations with respect to the
Christoffel symbols $D^m{}_{ik}$. From now on, to avoid
considerable technical problems, restrict ourselves by such metrics
that for any basis vectors $E_i,E_k$ the values of $g_{ik}$ are
{\em real}, therefore

\begin{equation}
{\rm Im}g_{ik} = 0 \Rightarrow g_{ik} = g_{ki}
\label{img0}
\end{equation}

\noindent Now rewrite the first equation (\ref{c}) using (\ref{t0}):

\[
D^m{}_{ik}g_{ml} + D^m{}_{li}g_{km} = 2L^m{}_{li}g_{km}
\]

\noindent and perform the cyclic permutation of the free indices
$(ikl)$ taking (\ref{t0}) and (\ref{img0}) into account:

\begin{equation}
D^m{}_{ik}g_{lm} = L^m{}_{ik}g_{lm} +
L^m{}_{li}g_{km} - L^m{}_{kl}g_{im}
\label{dikj}
\end{equation}

\noindent thus, by virtue of (\ref{ndg}), the values of
$D^m{}_{ik}$ are completely determined by those of $g_{ik}$.
Therefore the remaining equations from (\ref{c}) are to be
considered as concertedness condition for $g^m{}_{ik}$ as well,
completely determining the values of the "non-commutative part" of
the metric when its commutative part $g_{ik}$ is given.

\paragraph{Remark.} I do not consider here a possible reasonable
weakening of the condition (\ref{ndg}), namely, the requirement of
the maximality of the rank of the matrix $(g_{ik},g^m{}_{ik})$. In
this case the role of $g_{lm}$ in (\ref{dikj}) can be played by a
non-degenerate submatrix of $(g_{ik},g^m{}_{ik})$.

\noindent Under the assumptions (\ref{ndg},\ref{img0}) the standard
convention on raising and lowering tensor indices by means of
$g_{ik}$ can be used. Thus (\ref{dikj}) can be rewritten as:

\begin{equation}
D_{lik} = L_{lik} + L_{kli} - L_{ikl}
\label{dlik}
\end{equation}

So, the formulas (\ref{dikj},\ref{dlik}) allow us to calculate the
coefficients of the connection concerted with the metric $g$.

\subsection{Einstein -- Hilbert variational principle.}

Having the nondegenerate metric tensor $g_{ik}$ in our disposal, we
can form the {\sc scalar curvature} as follows:

\begin{equation}
R = g^{ik}R_{ik} =
g^{ik}D^m{}_{ni}D^n{}_{nk} -
g^{ik}D^n{}_{nm}D^m{}_{ik}
\label{r0}
\end{equation}

\noindent where $g^{ik}$ denotes, as usually, the matrix inverse
to $g_{ik}$. Begin the analysis of this formula from its second
term $g^{ik}D^n{}_{nm}D^m{}_{ik}$.  It contains the factor
$D^n{}_{nm}$, which, according to

\[D^n{}_{nm} = g^{ik}D_{ikm} = 2L^k{}_{km} = 2L_m\]

\noindent does not depend on the metric. So, the second term of
(\ref{r0}) reads:

\begin{equation}
g^{ik}D^n{}_{nm}D^m{}_{ik} = 2g^{ik}L_{m}D^m{}_{ik} =
g^{ik}L^lD_{lik} = -4g^{ik}L_iL_k
\label{2t}
\end{equation}

\noindent The routine manipulations with indices turn the first
term of (\ref{r0}) to:

\[ g^{ik}D^m{}_{ni}D^n{}_{mi} = D_{mni}D^{nmi} =
g^{ik}g^{jl}g_{mn}L^m{}_{ij}L^n{}_{kl} + 2g^{ik}K_{ik} \]

\noindent So, (\ref{r0}) reads:

\begin{equation}
R = g^{ik}(g^{jl}g_{mn}L^m{}_{ij}L^n{}_{kl} + 2K_{ik} + 4L_iL_k)
\label{r}
\end{equation}

\noindent Note that $R$ has already numeric values rather
than scalar ones, so we can consider its variation:

\[ \delta R = 2\delta g^{ik}(L{min}L^m{}_k{}^n + K_{ik} + 2L_iL_k -
\frac{1}{2}L_{imn}L_k{}^{mn}) = 2G_{ik}\delta g^{ik} \]

\noindent where $G_{ik}$ is the analog of the {\sc Einstein
tensor}:

\begin{equation}
G_{ik} = L_{min}L^m{}_k{}^n - \frac{1}{2}L_{imn}L_k{}^{mn} + K_{ik}
+ 2L_iL_k
\label{ein}
\end{equation}

\noindent Note that the first two terms of (\ref{ein}) depend on the
metric unlike the remaining two: $K_{ik}$ and $L_iL_k$.

\subsection{Eigen-subalgebras and topologimeter.}

Suppose we managed to solve the finitary matrix analog of the
Einstein equation

\begin{equation}
G=T
\label{ee}
\end{equation}

\noindent where $G$ is the above defined Einstein tensor
(\ref{ein}) and $T$ is a finitary counterpart of the stress-energy
tensor. And suppose that the resulting metric tensor $g_{ik}$
splits the linear space $V$ into a set of mutually orthogonal (with
respect to $g$) subspaces:

\begin{equation}
V = V_1 + V_2 + \ldots
\label{vdec}
\end{equation}

The crucial point of the techniques suggested is that we may
consider the elements of $V$ as those of the basic algebra ${\cal
S}$.  Therefore the decomposition (\ref{vdec}) gives rise to the
following decomposition of ${\cal S}$:

\begin{equation}
{\cal S} = E_0 + {\cal A}_1 + {\cal A}_2 + \ldots + \hbox{ perhaps, some
remainder}
\label{adec}
\end{equation}

\noindent that is, with each $V_i$ from (\ref{vdec}) we can associate a
subalgebra ${\cal A}_i\subseteq {\cal S}$ spanned on {\em the linear
subspace} $V_i$ of ${\cal S}$. Consequently, we can apply the
spatialization procedure described in section \ref{spat} to each
${\cal A}_i$, and then the solution $g_{ik}$ of the Einstein-like
equation (\ref{ee}) produces the family $\{X_i\}$ of pattern spaces
(\ref{dspat}):

\begin{equation}
X_i = {\rm spat}({\rm span}_{\cal A}(V_i))
\label{xi}
\end{equation}

In standard quantum mechanics an apparatus measuring a entity $Q$
can be described as follows. We have the state space ${\cal H}$ of
a system and a family of mutually orthogonal (with respect to the
inner product in ${\cal H}$) subspaces $\{{\cal H}_i\subseteq {\cal
H}\}$. With each subspace ${\cal H}_i$ we associate a value of the
measured entity $Q$. In standard quantum mechanics these values are
real numbers, though it is a mere matter of choice.

Now return to the suggested machinery. We can think of the pair
$(V,g)$ as state space, and consider the decomposition (\ref{vdec})
as that associated with the measuring apparatus. But what should we
assign to each $V_i$? The answer is given by (\ref{xi}): these are
pattern spaces. So, we may conclude that the finite-topology-valued
observable on the state space $V$ is built, and the hypothetical
device associated with the partition (\ref{vdec}) may thus be
called {\sc topologimeter}.

\medskip
\section{Concluding remarks.}

The machinery was suggested to draw the idea of description of
varying spacetime topology to the level of calculations. It
consists of the following:

\begin{itemize}
\item {\bf The coarse-graining procedure} which replaces the
continuous spacetime by finitary pattern space described in section
\ref{1}.

\item {\bf The incidence algebras} associated with pattern spaces
were introduced in section \ref{2} to replace graphs being discrete
object by their nature by linear spaces making it possible to
embody them in a greater object (algebraic superspace) of the same
type and enable the possibility to describe continuous evolution.

\item {\bf The spatialization procedure} suggested in section
\ref{spat} is the inverse to the construction of incidence
algebras and produces pattern spaces by given
fi\-ni\-te-di\-men\-si\-o\-nal algebras.

\item {\bf The Einstein -- Hilbert variational principle} in
finitary form was written down in section \ref{5} for the algebraic
superspace.  It can give rise to a family of mutually orthogonal
subspaces each of which is associated (via spatialization
procedure) with certain topology. The construction is interpreted
as the mathematical description of {\em topologimeter}.

\end{itemize}

So, what are the consequences of the suggested approach? We now
have the machinery in our disposal which is able to describe the
changing spacetime topology on the level of pattern spaces.
It is seen within this approach that it is pointless to
speak of both events forming the spacetime and its topology before
a particular measurement is performed. Moreover, the situation when
we can speak of separation of events looks very special, namely,
the solution of the Einstein equation in the superspace must
support the decomposition (\ref{vdec}) which may not take place.

I should also mention a crucial question remaining beyond the scope
of the presented work. It is the correspondence principle: {\em to
what extent pattern spaces really substitute continuous spacetime}.
There are two modes of answering this question. That first is to
claim that, as a matter of fact, nobody is able to prove that
spacetime is really continuous: there is no operationalistaically
sound procedure checking the continuity of the spacetime since
it would require to consider the infinite number of events.
Moreover, the individual event itself is an idealization rather
than a testable entity \cite{geroch1}. Another way to corroborate
the correspondence principle is to use the techniques proposed in
\cite{sorkin} where the inverse limits of pattern spaces converging
to continuous manifolds are studied.

The finite-dimensional models I consider may be applied to other
fundamental theories which are devoted to describe the structure of
spacetime. I will dwell upon two such theories. The first is the
histories approach to quantum mechanics suggested by Griffiths
and Hartle \cite{ha}. In Isham's \cite{ishql} version of this
approach each particular pattern space (substituting a spacetime)
may serve as the counterpart of a particular history, and the
application of topologimeter can be associated with the
decoherence if we assign a numeric value to each subspace of the
decomposition (\ref{vdec}). The second theory where the machinery I
suggest may be relevant is the construction of spacetime from
elementary consituents called {\sc urs} \cite{gornitz}. When we
replace spacetime by a pattern space, we may consider the latter as
the set of linked darts and then ask what is the law linking them.
Each dart, in turn, can be described by the smallest pattern space
consisting of two points: the appropriate algebraic superspace for
each virtual dart is the algebra of $2\times 2$ complex matrices,
where the Einstein like equations (\ref{ein}) can be solved
completely without any additional requirements.

\paragraph{Acknowledgments.} The author appreciates the attention
offered to the work by the participants of the {\em Quantum
Structures'94} Meeting held in Prague in August, 1994 and much
helpful advice from A.A. Grib, S.V. Krasnikov and A.A. Lobashov.
The financial support from the ISF (George Soros Emergency Grant)
is acknowledged.

\end{document}